\documentclass[reprint,floatfix]{revtex4-1}

\usepackage[frenchb]{babel}
\usepackage[T1]{fontenc}
\usepackage{graphicx}
\usepackage{natbib} 
\usepackage{amsmath}
\usepackage{amssymb}
\usepackage{color}

\begin{document}

\renewcommand\refname{}

\title{Magnetometry of micro-magnets with electrostatically defined Hall bars}

\author{Dany~\surname{Lachance-Quirion}}
\affiliation{Département de Physique, Université de Sherbrooke, J1K 2R1, Sherbrooke, Québec, Canada}

\author{Julien~\surname{Camirand Lemyre}}
\affiliation{Département de Physique, Université de Sherbrooke, J1K 2R1, Sherbrooke, Québec, Canada}

\author{Laurent~\surname{Bergeron}}
\affiliation{Département de Physique, Université de Sherbrooke, J1K 2R1, Sherbrooke, Québec, Canada}

\author{Christian~\surname{Sarra-Bournet}}
\affiliation{Département de Physique, Université de Sherbrooke, J1K 2R1, Sherbrooke, Québec, Canada}

\author{Michel~\surname{Pioro-Ladrière}}
\email{michel.pioro-ladriere@usherbrooke.ca}
\affiliation{Département de Physique, Université de Sherbrooke, J1K 2R1, Sherbrooke, Québec, Canada}
\affiliation{CIFAR Program in Quantum Information Science, Canadian Institute for Advanced Research (CIFAR), M5G 1Z8, Toronto, Ontario, Canada}

\date{December 6 2015}

\begin{abstract}
\noindent Micro-magnets are key components for quantum information processing with individual spins, enabling arbitrary rotations and addressability. In this work, characterization of sub-micrometer sized CoFe ferromagnets is performed with Hall bars electrostatically defined in a two-dimensional electron gas. Due to the ballistic nature of electron transport in the cross junction of the Hall bar, anomalies such as the quenched Hall effect appear near zero external magnetic field, thus hindering the sensitivity of the magnetometer to small magnetic fields. However, it is shown that the sensitivity of the diffusive limit can be almost completely restored at low temperatures using a large current density in the Hall bar of about 10~A/m. Overcoming the size limitation of conventional etched Hall bars with electrostatic gating enables the measurement of magnetization curves of 440~nm wide micro-magnets with a signal-to-noise ratio above 10$^3$. Furthermore, the inhomogeneity of the stray magnetic field created by the micro-magnets is directly measured using the gate-voltage-dependent width of the sensitive area of the Hall bar.
\end{abstract}

\maketitle
Exciting progress towards quantum technologies has recently been made with electron spins in quantum dots~\cite{Simmons2007,Veldhorst2014,Veldhorst2015,Viennot2015}. The stray magnetic field of proximal ferromagnets enables fast single-qubit operations through electric-dipole spin resonance and addressability between neighbouring spins~\cite{Tokura2006,Laird2007,Pioro-Ladriere2008,Kawakami2014a,Yoneda2014,Forster2015}. However, for scaling up to multiple dot architectures, ferromagnets of sizes comparable to the lithographic dimensions of the dots, which can be as small as 50~nm~\cite{Lim2009a}, are desired. Magnetization properties of sub-micrometer-scale magnets depend greatly on the interplay between magnetocrystalline and shape anisotropies~\cite{Bonet1999}. Therefore, it is necessary to verify that individual ferromagnets produce large inhomogeneous magnetic fields at a low saturation field, two features desired for spin manipulation in quantum dots.\par
Magnetometry of individual magnets can be performed using techniques such as magnetic force microscopy~\cite{Gibson1991,Viennot2015}, micro-SQUID magnetometry~\cite{Wernsdorfer1996} and Hall magnetometry~\cite{Kent1994,Geim1997,Monzon1997}. In the latter, an in-plane magnetic field polarizes a ferromagnet placed close to a Hall bar, which creates a stray magnetic field in the Hall bar~\cite{Monzon1997}. A Hall voltage proportional to the out-of-plane component of the stray field averaged over the cross junction of the Hall bar is then created~\cite{Geim1997}. This Hall voltage is significant only when the Hall bar is carefully aligned and comparable in size with the ferromagnet. Micrometer-sized Hall bars can be fabricated by etching an heterostructure with a two dimensional electron gas (2DEG)~\cite{Kent1994,Geim1997,Monzon1997}. However, a large depletion zone at the edge of the etched area usually restricts the lateral dimensions of etched Hall bars to a micrometer~\cite{VanHouten1986,Choi1987,Heitmann1990}, limiting the signal-to-noise ratio for nanometer-scale ferromagnets~\cite{Kent1994,Monzon1999,Grundler1999,Meier2000,Schuh2001}. While sensitive area of Hall bars can be reduced to sub-micrometer dimensions with electrostatic gating~\cite{Thornton1986,Zheng1986,Ford1988}, a detailed study of these devices as magnetometers is lacking.\par
In this Letter, we present magnetometry results of individual sub-micrometer sized ferromagnets obtained using Hall bars defined electrostatically by depletion gates. The response of the magnetometer is first characterized in an external perpendicular magnetic field. Magnetization curves with high signal-to-noise ratios of two ferromagnet geometries are then presented. The electrostatic control over the active area of the Hall bar reveals the inhomogeneity of the stray magnetic field, a characteristic desired for spin manipulation with micro-magnets and not directly accessible with Hall bars of fixed dimensions such as etched bars. Moreover, electrostatic Hall bars can be incorporated into the fabrication of lateral quantum dot devices without additional steps, allowing on-chip micromagnetic characterization.\par
\begin{figure}
\centering
\includegraphics*[width = 1.00\columnwidth]{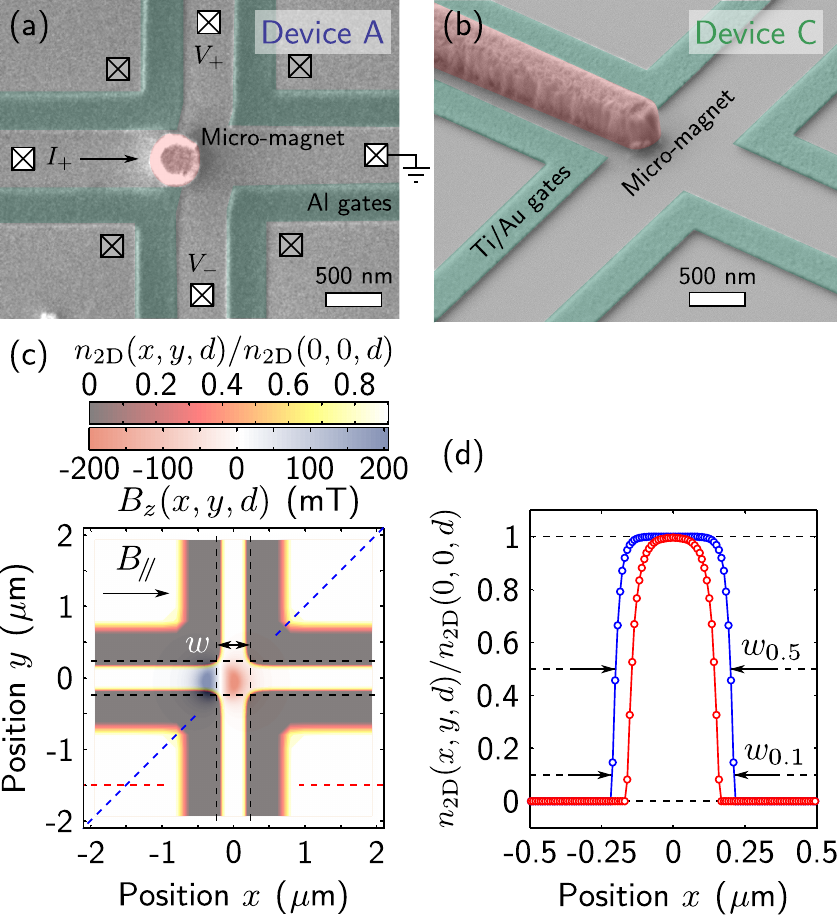}
\caption{False-colored SEM images of gated Hall bars with (a)~self-insulating aluminium depletion gates and 550~nm wide channels for devices A and B and (b)~titanium/gold depletion gates and 750~nm wide channels for device C. Both type of devices have a 300~nm-thick CoFe micro-magnet near the cross junction of the Hall bar. Ohmic contacts inside and outside the Hall bar are respectively shown as white and gray crossed squares in (a). The self-insulating property of aluminium gates can be used to avoid electrical contacts between the gates and the micro-magnet. (c) Combined simulations of the electron density $n_\text{2D}(x,y,d)$ profile for a gate voltage $V_g=-0.6$~V and of the magnetic field profile $B_z(x,y,d)$ of the micro-magnet of device A when saturated by an in-plane magnetic field $B_{\mathbin{\!/\mkern-5mu/\!}}$, both evaluated in the 2DEG ($d=100$~nm). Black dashed lines delimit the considered cross junction of width $w$ and blue and red dashed lines indicate lines along which density profiles of (d) are plotted. (d) Normalized density profiles $n_\text{2D}(x,y,d)/n_\text{2D}(0,0,d)$ along $x=y$ (blue, projected along $x$) and $y=-1.5~\mu$m (red) show different effective widths for the cross junction and the leads of the gated Hall bar. Widths calculated from the $x=y$ profile at half-maximum ($w_{0.5}$) and at 10\% of the maximum ($w_{0.1}$) are later considered.}
\label{Figure_1}
\end{figure}
Gated Hall bars are fabricated from a AlGaAs/GaAs heterostructure in which a 2DEG with an electron density of $n_\text{2D}=2.2\times10^{11}~\text{cm}^{-2}$ and a mobility of $\mu=1.7\times10^6~\text{cm}^2/(\text{V}\times\text{s})$ is formed at a distance~$d=100$~nm from the surface~\cite{Bureau-Oxton2013}. As shown in the scanning electron microscope (SEM) images of Fig.~\ref{Figure_1}~(a) and (b), depletion gates are shaped in order to define in the 2DEG a Hall bar which is only slightly larger than the magnets. Transport measurements show that ohmic contacts inside and outside the Hall bar are well isolated from each other for gate voltages below -0.55~V, indicating the formation of a Hall bar. Micro-magnets are fabricated using a standard lift-off process with electron-beam lithography followed by electron-beam deposition of 300~nm thick CoFe. Devices A and B each have a lithographically-identical cylindrical-shaped micro-magnet while a stadium-shaped magnet is present on device C.\par
A simulation of the electron density $n_\text{2D}$, performed using nextnano~\cite{Birner2007} for a gate voltage $V_g$ of -0.6~V, displays a well defined Hall bar in the 2DEG (Fig.~\ref{Figure_1}~(c)). The width $w$ of the cross junction of the Hall bar, the sensitive area of the magnetometer, cannot be determined accurately by transport measurements as current is first pinched off in the leads. Indeed, the cross section of the density profile along a diagonal, projected into the $x$ axis, shows a width significantly larger than along current-carrying leads (Fig.~\ref{Figure_1}~(d)).\par
The out-of-plane stray magnetic field $B_z(x,y,d)$, shown in Fig.~\ref{Figure_1}~(c) for device A, is simulated at saturation in a parallel field using Radia~\footnote{Mathematica Radia package available at http://www.esrf.eu/ is used for the simulations of the stray magnetic field.}. The saturated magnetization of 1.93~T is determined from measurements on CoFe thin films using a SQUID magnetometer. In order to maximize the detected magnetic field, the relative position between the Hall bar and the micro-magnet is chosen such that the magnetic field of a single pole of the magnet enters the cross junction of the Hall bar. With an optimal position of the micro-magnet, the average transverse magnetic field in the cross junction, $\langle B_z^\text{sat}\rangle$, reaches approximately 84~mT according to simulations, about 37\% of the peak magnetic field in the 2DEG.\par 
\begin{figure*}
\centering
\includegraphics*[width = 2.00\columnwidth]{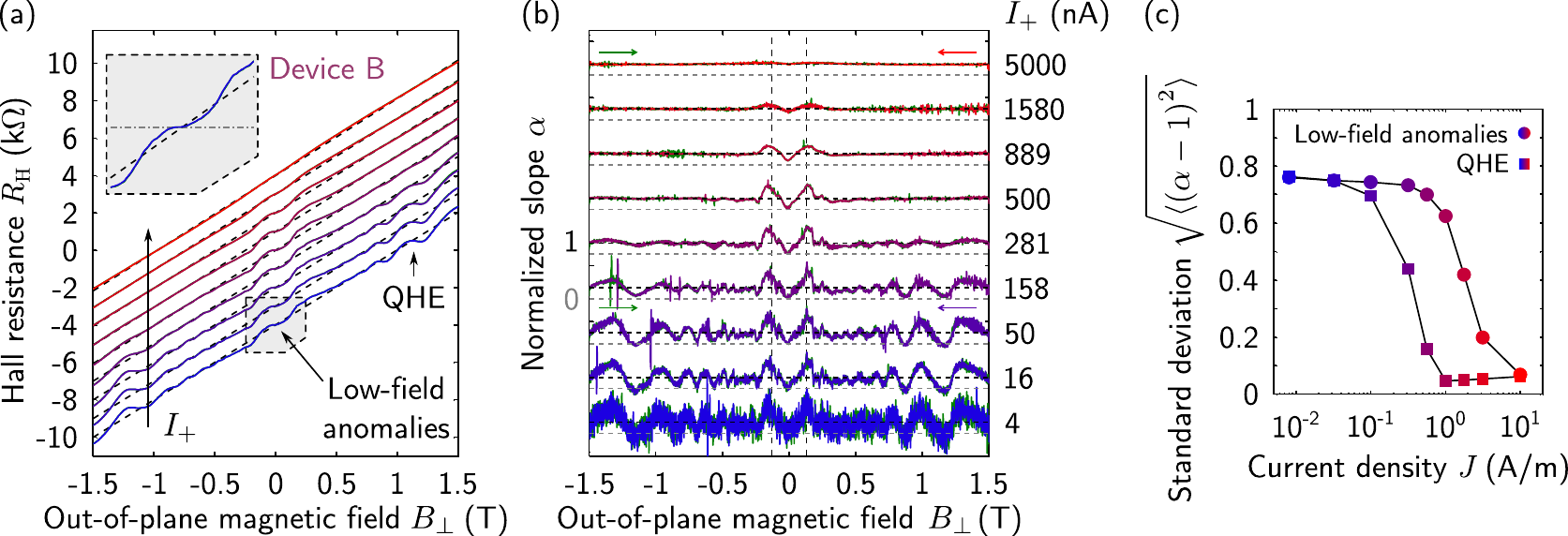}
\caption{(a) Hall resistance~$R_\text{H}$ as a function of an out-of-plane magnetic field~$B_\perp$ for device B for the different injection currents~$I_+$ indicated in (b) and for $V_g=-0.6$~V. Traces are offset by multiples of 1~k$\Omega$ for clarity and linear regressions are shown as black dashed lines, corresponding to an electron density~$n_\text{2D}=1.54\times10^{11}$~cm$^{-2}$. Inset: zoom of the anomaly near zero magnetic field for $I_+=4$~nA. (b) Numerical derivative of the Hall resistance measurements normalized by the slope of the corresponding linear regression, defined as $\alpha$. Positive to negative magnetic field sweeps are color-coded blue to red depending on the injection current while negative to positive sweeps are shown in green for all injection currents shown on the right axis. Black vertical dashed lines show the characteristic magnetic field~$\pm B_0=\pm130$~mT. Black and grey horizontal dashed lines show $\alpha=1$ and $\alpha=0$ respectively for each measurement. (c) Standard deviation of $\alpha$ from 1 in the magnetic field range~$\left|B_\perp\right|\leq B_0/2$, corresponding to low-field anomalies (circles), and $\left|B_\perp\right|\geq1$~T, corresponding to the QHE (squares), as a function of current density~$J=I_+/w$ with $w=500$~nm. Black lines are guides to the eye.}
\label{Figure_2}
\end{figure*}
A standard lock-in technique is used to measure the Hall voltage~$V_\text{H}\equiv V_+-V_-$ in phase with a low-frequency ac injection current~$I_+$ at a temperature of 1.5~K. Figure~\ref{Figure_2}~(a) shows the measurements of the Hall resistance~$R_\text{H}=V_\text{H}/I_+$ in a perpendicular magnetic field $B_\perp$ for a fixed gate voltage of $-0.6$~V applied to all gates of device B. The electron density in the cross junction~$n_\text{2D}=1.54\times10^{11}$~cm$^{-2}$, obtained by a linear regression given by $R_\text{H}(B_\perp)=B_\perp/en_\text{2D}$, is found to vary by less than 1\% when changing $I_+$ by three orders of magnitude.\par
Deviations from the $R_\text{H}\propto B_\perp$ behaviour are observed for injection currents below 500~nA with two distinct features. The plateaus at high fields ($B_\perp>0.75$~T) are due to the usual quantum Hall effect (QHE)~\cite{Klitzing1980}. The low-field anomalies are related to the \textit{last Hall plateau} and the \textit{quenched Hall effect}~\cite{Roukes1987}. Deviations from linearity are highlighted in Fig.~\ref{Figure_2}~(b) by plotting the normalized slope~$\alpha\equiv en_\text{2D}\left(\text{d}R_\text{H}/\text{d}B_\perp\right)$, calculated from the numerical derivative of $R_\text{H}$ normalized by the slope $1/en_\text{2D}$~\cite{Geim1997}. A near-zero $\alpha$ implies that the Hall magnetometer is insensitive to stray magnetic fields.\par
Low-field anomalies require ballistic electron transport and rounded corners~\cite{Beenakker1990}, which is the case for our gated Hall bars (Fig.~\ref{Figure_1}~(c)). These anomalies should be visible when $B_\perp$ is smaller than a characteristic magnetic field~$B_0=\hbar k_F/ew$, where $k_F=\sqrt{2\pi n_\text{2D}}$ is the Fermi wavevector~\cite{Beenakker1989}. Considering a cross junction width~$w$ of $500$~nm and the measured $n_\text{2D}$ for device B, the characteristic magnetic field~$B_0$ is 130~mT, which corresponds approximately to the observed range of deviations in Fig.~\ref{Figure_2}~(b).\par
\begin{figure}
\centering
\includegraphics*[width = 1.00\columnwidth]{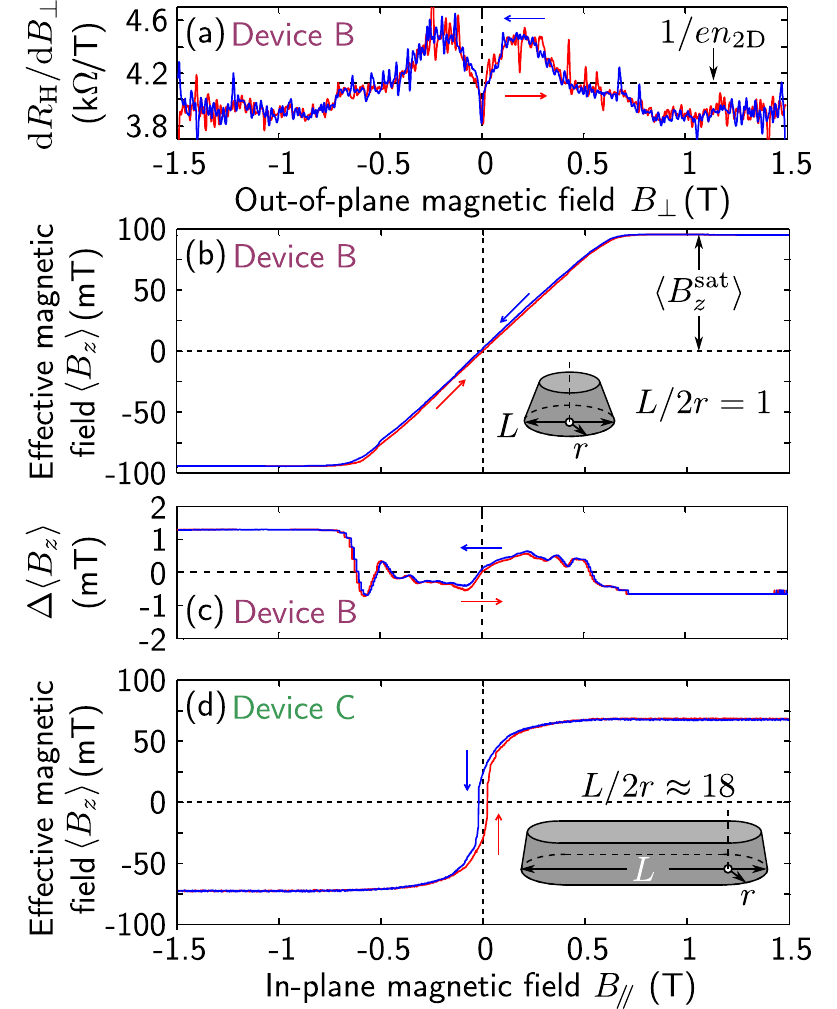}
\caption{(a) Derivative of the Hall resistance as a function of $B_\perp$ for device B with $I_+=5$~$\mu$A. (b) Effective magnetic field $\langle B_z\rangle$ of the micro-magnet as a function of an in-plane magnetic field $B_{\mathbin{\!/\mkern-5mu/\!}}$ for the cylindrical micro-magnet ($L/2r=1$, $r=230$~nm) of device B for $I_+=5$~$\mu$A. (c) Difference $\Delta\langle B_z\rangle$ between the corrected effective magnetic field $\langle B_z\rangle$ and the effective magnetic field calculated using $\langle B_z\rangle'=en_\text{2D}R_\text{H}(B_{\mathbin{\!/\mkern-5mu/\!}})$ for the magnetization curve of device B shown in (b). (d)~Magnetization curve of the 8~$\mu$m long stadium-shaped micro-magnet ($L/2r\approx18$, $r=220$~nm) of device C with $I_+=2$~$\mu$A. All these measurements are performed using a gate voltage $V_g=-0.6$~V applied on all gates. Blue and red lines respectively indicate positive to negative and negative to positive magnetic field sweeps.}
\label{Figure_3}
\end{figure}

Both the QHE and the low-field anomalies can be quenched at a temperature of 1.5~K using a high current density $J\equiv I_+/w$. Indeed, as shown in Fig.~\ref{Figure_2}~(c), the standard deviation of $\alpha$ from 1 at high fields indicates a critical current density of about 1~A/m for the QHE~\cite{Ebert1983} while a current density ten times larger is necessary to almost completely quench low-field anomalies. The Hall resistance deviates only slightly from its linear behaviour at an injection current of about 5~$\mu$A (Fig.~\ref{Figure_3}~(a)). While higher current densities could be used to further suppress the anomalies, we find that the Hall effect becomes again nonlinear for current densities above a threshold depending on the sample, gate voltage and external magnetic field.\par
In a external in-plane magnetic field $B_{\mathbin{\!/\mkern-5mu/\!}}$, the ferromagnets produce a stray magnetic field with an out-of-plane component at the level of the 2DEG. The magnetization curves of the ferromagnets are then obtained by measuring the Hall resistance $R_\text{H}$ as a function of $B_{\mathbin{\!/\mkern-5mu/\!}}$~\cite{Monzon1997}. The small nonlinearity of $R_\text{H}(B_\perp)$ remaining at high current densities can be accounted for when evaluating the effective magnetic field~$\langle B_z\rangle$ created by the micro-magnet. The normalized response $\alpha$ of the magnetometer is considered to be given by $\alpha(B_\perp)$ measured in a perpendicular magnetic field $B_\perp$ but evaluated at the effective magnetic field $\langle B_z\rangle$, such that
\begin{equation}
\langle B_z\rangle(B_{\mathbin{\!/\mkern-5mu/\!}})=\frac{en_\text{2D}R_\text{H}(B_{\mathbin{\!/\mkern-5mu/\!}})}{\alpha(\langle B_z\rangle)}.
\label{Equation_1}
\end{equation}
 An iterative method is used to solve this equation. The magnetic field $\langle B_z^{(i)}\rangle(B_{\mathbin{\!/\mkern-5mu/\!}})$  at the $i$th step is evaluated using $\alpha(\langle B_z^{(i-1)}\rangle)$. Starting from the initial condition $\alpha(\langle B_z^{(0)}\rangle)=1$, the process is repeated until convergence. Figure~\ref{Figure_3}~(b) and (d) show the magnetization curves of the micro-magnets of devices B and C corrected using this method. The signal-to-noise ratio of $\langle B_z\rangle$ at saturation is higher than $10^3$ for devices A and B and slightly above $10^2$ for device C~\footnote{SEM observation was performed on device C before measurements causing telegraphic noise and lowering the signal-to-noise ratio.}. The small discrepancy between corrected and uncorrected ($\alpha=1$) effective magnetic fields implies that a high current density is enough to obtain without further corrections reliable magnetization curves with gated Hall bars~(Fig.~\ref{Figure_3}~(c)).\par
\begin{figure}
\centering
\includegraphics*[width = 1.00\columnwidth]{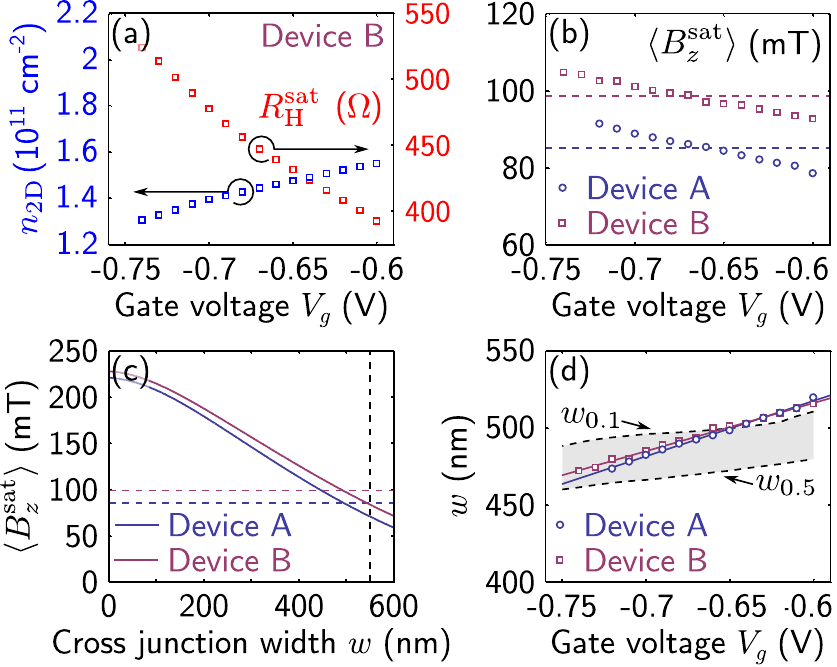}
\caption{(a) Electron density $n_\text{2D}$ (blue squares) and Hall resistance at saturation $R_\text{H}^\text{sat}$ (red squares) as a function of the gate voltage $V_g$ for device B. (b) Effective magnetic field at saturation $\langle B_z^\text{sat}\rangle$ as a function of gate voltage $V_g$ for devices A (blue circles) and B (purple squares). Dashed lines indicate gate-voltage averaged $\langle B_z^\text{sat}\rangle$ for both devices. (c) $\langle B_z^\text{sat}\rangle$ as a function of the cross junction width $w$ as calculated using equation~\eqref{Equation_2} and the simulated magnetic field profile $B_z(x,y,d)$ of the micro-magnet in the plane of the 2DEG ($d=100$~nm) for devices A (blue plain line) and B (purple plain line). The vertical dashed line shows the lithographic width of 550~nm of the devices while horizontal dashed lines are the same than in (b). (d) Cross junction width $w$ as a function of gate voltage $V_g$ as determined by comparing (b) and (c) for devices A (blue circles) and B (purple squares). Plain lines show linear fits to data for each device while dashed lines show the width extracted from electrostatic simulations of the electron density profile in the cross junction at half-maximum ($w_{0.5}$) and 10\% of the maximum ($w_{0.1}$) as shown in Fig.~\ref{Figure_1}~(d). In all data of this figure, error bars corresponding to three standard deviations of the measured quantities for $\left|B_{\mathbin{\!/\mkern-5mu/\!}}\right|>1$~T are smaller than the symbols.}
\label{Figure_4}
\end{figure}
The gate voltage dependence of the electron density $n_\text{2D}$ and the Hall resistance at saturation $R_\text{H}^\text{sat}$ is shown in Fig.~\ref{Figure_4}~(a) for device~B. The increase of the Hall resistance at saturation when the gate voltage $V_g$ becomes more negative cannot be entirely explained by a decreasing electron density. Because the stray magnetic field is highly inhomogeneous near the cross junction, the average magnetic field depends on the width of the cross junction in the 2DEG which in turns depends on the gate voltage. As shown in Fig.~\ref{Figure_4}~(b), the effective magnetic field at saturation $\langle B_z^\text{sat}\rangle$, calculated using Eq.~\eqref{Equation_1}, is found to increase for both devices A and B when decreasing gate voltage. This provides a direct measurement of the inhomogeneity of the stray magnetic field of the micro-magnets.\par
To be more quantitative, the measured effective transverse magnetic field can be related to the cross junction width $w$ using the simulated magnetic profile $B_z(x,y,d)$ according to
\begin{equation}
\langle B_z^\text{sat}\rangle(w)=\int_{-w/2}^{w/2}\int_{-w/2}^{w/2}\text{d}x\ \text{d}y\ B_z(x,y,d)/w^2,
\label{Equation_2}
\end{equation}
where the origin $(x=0,y=0)$ is defined at the center of the cross junction as in Fig.~\ref{Figure_1}~(c). Figure~\ref{Figure_4}~(c) shows for devices A and B the results of the calculation in which the relative position between the cross junction and the micro-magnet is determined from SEM images. The measurements $\langle B_z^\text{sat}\rangle(V_g)$ (Fig.~\ref{Figure_4}~(b)) are 
then compared to the calculations $\langle B_z^\text{sat}\rangle(w)$ (Fig.~\ref{Figure_4}~(c)) to determine the cross junction width $w$ corresponding to each gate voltage $V_g$. A linear dependence on gate voltage is observed (Fig.~\ref{Figure_4}~(d)), consistent with previous measurements in one-dimensional channels~\cite{Zheng1986}. Furthermore, the values of $w$ between lithographically identical devices A and B are very similar. This shows that the magnetic fields differences in Fig.~\ref{Figure_4}~(b) are mostly explained by the alignment of the micro-magnets, differing by approximately 40~nm according to SEM images. Finally, the cross junction widths calculated from electrostatic simulations are compared in Fig.~\ref{Figure_4}~(d) to the experimentally estimated widths. The good agreement with the simulations indicates that the measured effective magnetic field can be approximated by the average magnetic field in the cross junction of the Hall bar.\par
In conclusion, Hall bars electrostatically-defined in a two-dimensional electron gas have been used to perform magnetometry of sub-micrometer-sized ferromagnets. Electrostatic control over the active area of the Hall magnetometer has enabled measurements of the inhomogeneity of the stray magnetic field created by the micro-magnets. Low-field anomalies initially limiting the sensitivity of the magnetometer have been almost completely quenched using a large current density of about 10~A/m. The high signal-to-noise ratio of the measured magnetization curves makes gated Hall bars a sensitive and practical tool for the development of spin qubits with micro-magnets and for precise magnetometry of mesoscopic magnetic systems, such as nanometer-scale ferromagnets and superconductors.
\\\\
The authors would like to thank M. Lacerte for technical assistance and C. Bureau-Oxton for SQUID magnetometry of CoFe thin films. This work was supported by the Natural Sciences and Engineering Research Council of Canada (NSERC) and the Canada Foundation for Innovation (CFI).

\bibliographystyle{apsrev4-1}
\bibliography{Hall_magnetometry_APL2015bib}

\begin{thebibliography}{36}%
\makeatletter
\providecommand \@ifxundefined [1]{%
 \@ifx{#1\undefined}
}%
\providecommand \@ifnum [1]{%
 \ifnum #1\expandafter \@firstoftwo
 \else \expandafter \@secondoftwo
 \fi
}%
\providecommand \@ifx [1]{%
 \ifx #1\expandafter \@firstoftwo
 \else \expandafter \@secondoftwo
 \fi
}%
\providecommand \natexlab [1]{#1}%
\providecommand \enquote  [1]{``#1''}%
\providecommand \bibnamefont  [1]{#1}%
\providecommand \bibfnamefont [1]{#1}%
\providecommand \citenamefont [1]{#1}%
\providecommand \href@noop [0]{\@secondoftwo}%
\providecommand \href [0]{\begingroup \@sanitize@url \@href}%
\providecommand \@href[1]{\@@startlink{#1}\@@href}%
\providecommand \@@href[1]{\endgroup#1\@@endlink}%
\providecommand \@sanitize@url [0]{\catcode `\\12\catcode `\$12\catcode
  `\&12\catcode `\#12\catcode `\^12\catcode `\_12\catcode `\%12\relax}%
\providecommand \@@startlink[1]{}%
\providecommand \@@endlink[0]{}%
\providecommand \url  [0]{\begingroup\@sanitize@url \@url }%
\providecommand \@url [1]{\endgroup\@href {#1}{\urlprefix }}%
\providecommand \urlprefix  [0]{URL }%
\providecommand \Eprint [0]{\href }%
\providecommand \doibase [0]{http://dx.doi.org/}%
\providecommand \selectlanguage [0]{\@gobble}%
\providecommand \bibinfo  [0]{\@secondoftwo}%
\providecommand \bibfield  [0]{\@secondoftwo}%
\providecommand \translation [1]{[#1]}%
\providecommand \BibitemOpen [0]{}%
\providecommand \bibitemStop [0]{}%
\providecommand \bibitemNoStop [0]{.\EOS\space}%
\providecommand \EOS [0]{\spacefactor3000\relax}%
\providecommand \BibitemShut  [1]{\csname bibitem#1\endcsname}%
\let\auto@bib@innerbib\@empty
\bibitem [{\citenamefont {Simmons}\ \emph {et~al.}(2007)\citenamefont
  {Simmons}, \citenamefont {Thalakulam}, \citenamefont {Shaji}, \citenamefont
  {Klein}, \citenamefont {Qin}, \citenamefont {Blick}, \citenamefont {Savage},
  \citenamefont {Lagally}, \citenamefont {Coppersmith},\ and\ \citenamefont
  {Eriksson}}]{Simmons2007}%
  \BibitemOpen
  \bibfield  {author} {\bibinfo {author} {\bibfnamefont {C.~B.}\ \bibnamefont
  {Simmons}}, \bibinfo {author} {\bibfnamefont {M.}~\bibnamefont {Thalakulam}},
  \bibinfo {author} {\bibfnamefont {N.}~\bibnamefont {Shaji}}, \bibinfo
  {author} {\bibfnamefont {L.~J.}\ \bibnamefont {Klein}}, \bibinfo {author}
  {\bibfnamefont {H.}~\bibnamefont {Qin}}, \bibinfo {author} {\bibfnamefont
  {R.~H.}\ \bibnamefont {Blick}}, \bibinfo {author} {\bibfnamefont {D.~E.}\
  \bibnamefont {Savage}}, \bibinfo {author} {\bibfnamefont {M.~G.}\
  \bibnamefont {Lagally}}, \bibinfo {author} {\bibfnamefont {S.~N.}\
  \bibnamefont {Coppersmith}}, \ and\ \bibinfo {author} {\bibfnamefont {M.~A.}\
  \bibnamefont {Eriksson}},\ }\href {\doibase 10.1063/1.2816331} {\bibfield
  {journal} {\bibinfo  {journal} {Applied Physics Letters}\ }\textbf {\bibinfo
  {volume} {91}},\ \bibinfo {pages} {213103} (\bibinfo {year}
  {2007})}\BibitemShut {NoStop}%
\bibitem [{\citenamefont {Veldhorst}\ \emph {et~al.}(2014)\citenamefont
  {Veldhorst}, \citenamefont {Hwang}, \citenamefont {Yang}, \citenamefont
  {Leenstra}, \citenamefont {de~Ronde}, \citenamefont {Dehollain},
  \citenamefont {Muhonen}, \citenamefont {Hudson}, \citenamefont {Itoh},
  \citenamefont {Morello},\ and\ \citenamefont {Dzurak}}]{Veldhorst2014}%
  \BibitemOpen
  \bibfield  {author} {\bibinfo {author} {\bibfnamefont {M.}~\bibnamefont
  {Veldhorst}}, \bibinfo {author} {\bibfnamefont {J.~C.~C.}\ \bibnamefont
  {Hwang}}, \bibinfo {author} {\bibfnamefont {C.~H.}\ \bibnamefont {Yang}},
  \bibinfo {author} {\bibfnamefont {A.~W.}\ \bibnamefont {Leenstra}}, \bibinfo
  {author} {\bibfnamefont {B.}~\bibnamefont {de~Ronde}}, \bibinfo {author}
  {\bibfnamefont {J.~P.}\ \bibnamefont {Dehollain}}, \bibinfo {author}
  {\bibfnamefont {J.~T.}\ \bibnamefont {Muhonen}}, \bibinfo {author}
  {\bibfnamefont {F.~E.}\ \bibnamefont {Hudson}}, \bibinfo {author}
  {\bibfnamefont {K.~M.}\ \bibnamefont {Itoh}}, \bibinfo {author}
  {\bibfnamefont {A.}~\bibnamefont {Morello}}, \ and\ \bibinfo {author}
  {\bibfnamefont {A.~S.}\ \bibnamefont {Dzurak}},\ }\href {\doibase
  10.1038/nnano.2014.216} {\bibfield  {journal} {\bibinfo  {journal} {Nature
  Nanotechnology}\ }\textbf {\bibinfo {volume} {9}},\ \bibinfo {pages} {981}
  (\bibinfo {year} {2014})}\BibitemShut {NoStop}%
\bibitem [{\citenamefont {Veldhorst}\ \emph {et~al.}(2015)\citenamefont
  {Veldhorst}, \citenamefont {Yang}, \citenamefont {Hwang}, \citenamefont
  {Huang}, \citenamefont {Dehollain}, \citenamefont {Muhonen}, \citenamefont
  {Simmons}, \citenamefont {Laucht}, \citenamefont {Hudson}, \citenamefont
  {Itoh}, \citenamefont {Morello},\ and\ \citenamefont
  {Dzurak}}]{Veldhorst2015}%
  \BibitemOpen
  \bibfield  {author} {\bibinfo {author} {\bibfnamefont {M.}~\bibnamefont
  {Veldhorst}}, \bibinfo {author} {\bibfnamefont {C.~H.}\ \bibnamefont {Yang}},
  \bibinfo {author} {\bibfnamefont {J.~C.~C.}\ \bibnamefont {Hwang}}, \bibinfo
  {author} {\bibfnamefont {W.}~\bibnamefont {Huang}}, \bibinfo {author}
  {\bibfnamefont {J.~P.}\ \bibnamefont {Dehollain}}, \bibinfo {author}
  {\bibfnamefont {J.~T.}\ \bibnamefont {Muhonen}}, \bibinfo {author}
  {\bibfnamefont {S.}~\bibnamefont {Simmons}}, \bibinfo {author} {\bibfnamefont
  {A.}~\bibnamefont {Laucht}}, \bibinfo {author} {\bibfnamefont {F.~E.}\
  \bibnamefont {Hudson}}, \bibinfo {author} {\bibfnamefont {K.~M.}\
  \bibnamefont {Itoh}}, \bibinfo {author} {\bibfnamefont {A.}~\bibnamefont
  {Morello}}, \ and\ \bibinfo {author} {\bibfnamefont {A.~S.}\ \bibnamefont
  {Dzurak}},\ }\href {\doibase 10.1038/nature15263} {\bibfield  {journal}
  {\bibinfo  {journal} {Nature}\ }\textbf {\bibinfo {volume} {526}},\ \bibinfo
  {pages} {410} (\bibinfo {year} {2015})}\BibitemShut {NoStop}%
\bibitem [{\citenamefont {Viennot}\ \emph {et~al.}(2015)\citenamefont
  {Viennot}, \citenamefont {Dartiailh}, \citenamefont {Cottet},\ and\
  \citenamefont {Kontos}}]{Viennot2015}%
  \BibitemOpen
  \bibfield  {author} {\bibinfo {author} {\bibfnamefont {J.~J.}\ \bibnamefont
  {Viennot}}, \bibinfo {author} {\bibfnamefont {M.~C.}\ \bibnamefont
  {Dartiailh}}, \bibinfo {author} {\bibfnamefont {A.}~\bibnamefont {Cottet}}, \
  and\ \bibinfo {author} {\bibfnamefont {T.}~\bibnamefont {Kontos}},\ }\href
  {http://www.sciencemag.org/content/349/6246/408.abstract} {\bibfield
  {journal} {\bibinfo  {journal} {Science}\ }\textbf {\bibinfo {volume}
  {349}},\ \bibinfo {pages} {408} (\bibinfo {year} {2015})}\BibitemShut
  {NoStop}%
\bibitem [{\citenamefont {Tokura}\ \emph {et~al.}(2006)\citenamefont {Tokura},
  \citenamefont {van~der Wiel}, \citenamefont {Obata},\ and\ \citenamefont
  {Tarucha}}]{Tokura2006}%
  \BibitemOpen
  \bibfield  {author} {\bibinfo {author} {\bibfnamefont {Y.}~\bibnamefont
  {Tokura}}, \bibinfo {author} {\bibfnamefont {W.~G.}\ \bibnamefont {van~der
  Wiel}}, \bibinfo {author} {\bibfnamefont {T.}~\bibnamefont {Obata}}, \ and\
  \bibinfo {author} {\bibfnamefont {S.}~\bibnamefont {Tarucha}},\ }\href
  {\doibase 10.1103/PhysRevLett.96.047202} {\bibfield  {journal} {\bibinfo
  {journal} {Physical Review Letters}\ }\textbf {\bibinfo {volume} {96}},\
  \bibinfo {pages} {047202} (\bibinfo {year} {2006})}\BibitemShut {NoStop}%
\bibitem [{\citenamefont {Laird}\ \emph {et~al.}(2007)\citenamefont {Laird},
  \citenamefont {Barthel}, \citenamefont {Rashba}, \citenamefont {Marcus},
  \citenamefont {Hanson},\ and\ \citenamefont {Gossard}}]{Laird2007}%
  \BibitemOpen
  \bibfield  {author} {\bibinfo {author} {\bibfnamefont {E.~A.}\ \bibnamefont
  {Laird}}, \bibinfo {author} {\bibfnamefont {C.}~\bibnamefont {Barthel}},
  \bibinfo {author} {\bibfnamefont {E.~I.}\ \bibnamefont {Rashba}}, \bibinfo
  {author} {\bibfnamefont {C.~M.}\ \bibnamefont {Marcus}}, \bibinfo {author}
  {\bibfnamefont {M.~P.}\ \bibnamefont {Hanson}}, \ and\ \bibinfo {author}
  {\bibfnamefont {A.~C.}\ \bibnamefont {Gossard}},\ }\href {\doibase
  10.1103/PhysRevLett.99.246601} {\bibfield  {journal} {\bibinfo  {journal}
  {Physical Review Letters}\ }\textbf {\bibinfo {volume} {99}},\ \bibinfo
  {pages} {246601} (\bibinfo {year} {2007})}\BibitemShut {NoStop}%
\bibitem [{\citenamefont {Pioro-Ladri{\`{e}}re}\ \emph
  {et~al.}(2008)\citenamefont {Pioro-Ladri{\`{e}}re}, \citenamefont {Obata},
  \citenamefont {Tokura}, \citenamefont {Shin}, \citenamefont {Kubo},
  \citenamefont {Yoshida}, \citenamefont {Taniyama},\ and\ \citenamefont
  {Tarucha}}]{Pioro-Ladriere2008}%
  \BibitemOpen
  \bibfield  {author} {\bibinfo {author} {\bibfnamefont {M.}~\bibnamefont
  {Pioro-Ladri{\`{e}}re}}, \bibinfo {author} {\bibfnamefont {T.}~\bibnamefont
  {Obata}}, \bibinfo {author} {\bibfnamefont {Y.}~\bibnamefont {Tokura}},
  \bibinfo {author} {\bibfnamefont {Y.-S.}\ \bibnamefont {Shin}}, \bibinfo
  {author} {\bibfnamefont {T.}~\bibnamefont {Kubo}}, \bibinfo {author}
  {\bibfnamefont {K.}~\bibnamefont {Yoshida}}, \bibinfo {author} {\bibfnamefont
  {T.}~\bibnamefont {Taniyama}}, \ and\ \bibinfo {author} {\bibfnamefont
  {S.}~\bibnamefont {Tarucha}},\ }\href {\doibase 10.1038/nphys1053} {\bibfield
   {journal} {\bibinfo  {journal} {Nature Physics}\ }\textbf {\bibinfo {volume}
  {4}},\ \bibinfo {pages} {776} (\bibinfo {year} {2008})}\BibitemShut {NoStop}%
\bibitem [{\citenamefont {Kawakami}\ \emph {et~al.}(2014)\citenamefont
  {Kawakami}, \citenamefont {Scarlino}, \citenamefont {Ward}, \citenamefont
  {Braakman}, \citenamefont {Savage}, \citenamefont {Lagally}, \citenamefont
  {Friesen}, \citenamefont {Coppersmith}, \citenamefont {Eriksson},\ and\
  \citenamefont {Vandersypen}}]{Kawakami2014a}%
  \BibitemOpen
  \bibfield  {author} {\bibinfo {author} {\bibfnamefont {E.}~\bibnamefont
  {Kawakami}}, \bibinfo {author} {\bibfnamefont {P.}~\bibnamefont {Scarlino}},
  \bibinfo {author} {\bibfnamefont {D.~R.}\ \bibnamefont {Ward}}, \bibinfo
  {author} {\bibfnamefont {F.~R.}\ \bibnamefont {Braakman}}, \bibinfo {author}
  {\bibfnamefont {D.~E.}\ \bibnamefont {Savage}}, \bibinfo {author}
  {\bibfnamefont {M.~G.}\ \bibnamefont {Lagally}}, \bibinfo {author}
  {\bibfnamefont {M.}~\bibnamefont {Friesen}}, \bibinfo {author} {\bibfnamefont
  {S.~N.}\ \bibnamefont {Coppersmith}}, \bibinfo {author} {\bibfnamefont
  {M.~A.}\ \bibnamefont {Eriksson}}, \ and\ \bibinfo {author} {\bibfnamefont
  {L.~M.~K.}\ \bibnamefont {Vandersypen}},\ }\href {\doibase
  10.1038/nnano.2014.153} {\bibfield  {journal} {\bibinfo  {journal} {Nature
  Nanotechnology}\ }\textbf {\bibinfo {volume} {9}},\ \bibinfo {pages} {666}
  (\bibinfo {year} {2014})}\BibitemShut {NoStop}%
\bibitem [{\citenamefont {Yoneda}\ \emph {et~al.}(2014)\citenamefont {Yoneda},
  \citenamefont {Otsuka}, \citenamefont {Nakajima}, \citenamefont {Takakura},
  \citenamefont {Obata}, \citenamefont {Pioro-Ladri{\`{e}}re}, \citenamefont
  {Lu}, \citenamefont {Palmstr{\o}m}, \citenamefont {Gossard},\ and\
  \citenamefont {Tarucha}}]{Yoneda2014}%
  \BibitemOpen
  \bibfield  {author} {\bibinfo {author} {\bibfnamefont {J.}~\bibnamefont
  {Yoneda}}, \bibinfo {author} {\bibfnamefont {T.}~\bibnamefont {Otsuka}},
  \bibinfo {author} {\bibfnamefont {T.}~\bibnamefont {Nakajima}}, \bibinfo
  {author} {\bibfnamefont {T.}~\bibnamefont {Takakura}}, \bibinfo {author}
  {\bibfnamefont {T.}~\bibnamefont {Obata}}, \bibinfo {author} {\bibfnamefont
  {M.}~\bibnamefont {Pioro-Ladri{\`{e}}re}}, \bibinfo {author} {\bibfnamefont
  {H.}~\bibnamefont {Lu}}, \bibinfo {author} {\bibfnamefont {C.~J.}\
  \bibnamefont {Palmstr{\o}m}}, \bibinfo {author} {\bibfnamefont {A.~C.}\
  \bibnamefont {Gossard}}, \ and\ \bibinfo {author} {\bibfnamefont
  {S.}~\bibnamefont {Tarucha}},\ }\href {\doibase
  10.1103/PhysRevLett.113.267601} {\bibfield  {journal} {\bibinfo  {journal}
  {Physical Review Letters}\ }\textbf {\bibinfo {volume} {113}},\ \bibinfo
  {pages} {267601} (\bibinfo {year} {2014})}\BibitemShut {NoStop}%
\bibitem [{\citenamefont {Forster}\ \emph {et~al.}(2015)\citenamefont
  {Forster}, \citenamefont {M{\"{u}}hlbacher}, \citenamefont {Schuh},
  \citenamefont {Wegscheider},\ and\ \citenamefont {Ludwig}}]{Forster2015}%
  \BibitemOpen
  \bibfield  {author} {\bibinfo {author} {\bibfnamefont {F.}~\bibnamefont
  {Forster}}, \bibinfo {author} {\bibfnamefont {M.}~\bibnamefont
  {M{\"{u}}hlbacher}}, \bibinfo {author} {\bibfnamefont {D.}~\bibnamefont
  {Schuh}}, \bibinfo {author} {\bibfnamefont {W.}~\bibnamefont {Wegscheider}},
  \ and\ \bibinfo {author} {\bibfnamefont {S.}~\bibnamefont {Ludwig}},\ }\href
  {\doibase 10.1103/PhysRevB.91.195417} {\bibfield  {journal} {\bibinfo
  {journal} {Physical Review B}\ }\textbf {\bibinfo {volume} {91}},\ \bibinfo
  {pages} {195417} (\bibinfo {year} {2015})}\BibitemShut {NoStop}%
\bibitem [{\citenamefont {Lim}\ \emph {et~al.}(2009)\citenamefont {Lim},
  \citenamefont {Huebl}, \citenamefont {{Willems van Beveren}}, \citenamefont
  {Rubanov}, \citenamefont {Spizzirri}, \citenamefont {Angus}, \citenamefont
  {Clark},\ and\ \citenamefont {Dzurak}}]{Lim2009a}%
  \BibitemOpen
  \bibfield  {author} {\bibinfo {author} {\bibfnamefont {W.~H.}\ \bibnamefont
  {Lim}}, \bibinfo {author} {\bibfnamefont {H.}~\bibnamefont {Huebl}}, \bibinfo
  {author} {\bibfnamefont {L.~H.}\ \bibnamefont {{Willems van Beveren}}},
  \bibinfo {author} {\bibfnamefont {S.}~\bibnamefont {Rubanov}}, \bibinfo
  {author} {\bibfnamefont {P.~G.}\ \bibnamefont {Spizzirri}}, \bibinfo {author}
  {\bibfnamefont {S.~J.}\ \bibnamefont {Angus}}, \bibinfo {author}
  {\bibfnamefont {R.~G.}\ \bibnamefont {Clark}}, \ and\ \bibinfo {author}
  {\bibfnamefont {a.~S.}\ \bibnamefont {Dzurak}},\ }\href {\doibase
  10.1063/1.3124242} {\bibfield  {journal} {\bibinfo  {journal} {Applied
  Physics Letters}\ }\textbf {\bibinfo {volume} {94}},\ \bibinfo {pages}
  {173502} (\bibinfo {year} {2009})}\BibitemShut {NoStop}%
\bibitem [{\citenamefont {Bonet}\ \emph {et~al.}(1999)\citenamefont {Bonet},
  \citenamefont {Wernsdorfer}, \citenamefont {Barbara}, \citenamefont
  {Beno{\^{\i}}t}, \citenamefont {Mailly},\ and\ \citenamefont
  {Thiaville}}]{Bonet1999}%
  \BibitemOpen
  \bibfield  {author} {\bibinfo {author} {\bibfnamefont {E.}~\bibnamefont
  {Bonet}}, \bibinfo {author} {\bibfnamefont {W.}~\bibnamefont {Wernsdorfer}},
  \bibinfo {author} {\bibfnamefont {B.}~\bibnamefont {Barbara}}, \bibinfo
  {author} {\bibfnamefont {A.}~\bibnamefont {Beno{\^{\i}}t}}, \bibinfo {author}
  {\bibfnamefont {D.}~\bibnamefont {Mailly}}, \ and\ \bibinfo {author}
  {\bibfnamefont {A.}~\bibnamefont {Thiaville}},\ }\href {\doibase
  10.1103/PhysRevLett.83.4188} {\bibfield  {journal} {\bibinfo  {journal}
  {Physical Review Letters}\ }\textbf {\bibinfo {volume} {83}},\ \bibinfo
  {pages} {4188} (\bibinfo {year} {1999})}\BibitemShut {NoStop}%
\bibitem [{\citenamefont {Gibson}\ \emph {et~al.}(1991)\citenamefont {Gibson},
  \citenamefont {Smyth}, \citenamefont {Schultz},\ and\ \citenamefont
  {Kern}}]{Gibson1991}%
  \BibitemOpen
  \bibfield  {author} {\bibinfo {author} {\bibfnamefont {G.~A.}\ \bibnamefont
  {Gibson}}, \bibinfo {author} {\bibfnamefont {J.~F.}\ \bibnamefont {Smyth}},
  \bibinfo {author} {\bibfnamefont {S.}~\bibnamefont {Schultz}}, \ and\
  \bibinfo {author} {\bibfnamefont {D.~P.}\ \bibnamefont {Kern}},\ }\href@noop
  {} {\bibfield  {journal} {\bibinfo  {journal} {IEEE Transactions on
  Magnetics}\ }\textbf {\bibinfo {volume} {27}},\ \bibinfo {pages} {5187}
  (\bibinfo {year} {1991})}\BibitemShut {NoStop}%
\bibitem [{\citenamefont {Wernsdorfer}\ \emph {et~al.}(1996)\citenamefont
  {Wernsdorfer}, \citenamefont {Doudin}, \citenamefont {Mailly}, \citenamefont
  {Hasselbach}, \citenamefont {Benoit}, \citenamefont {Meier}, \citenamefont
  {Ansermet},\ and\ \citenamefont {Barbara}}]{Wernsdorfer1996}%
  \BibitemOpen
  \bibfield  {author} {\bibinfo {author} {\bibfnamefont {W.}~\bibnamefont
  {Wernsdorfer}}, \bibinfo {author} {\bibfnamefont {B.}~\bibnamefont {Doudin}},
  \bibinfo {author} {\bibfnamefont {D.}~\bibnamefont {Mailly}}, \bibinfo
  {author} {\bibfnamefont {K.}~\bibnamefont {Hasselbach}}, \bibinfo {author}
  {\bibfnamefont {a.}~\bibnamefont {Benoit}}, \bibinfo {author} {\bibfnamefont
  {J.}~\bibnamefont {Meier}}, \bibinfo {author} {\bibfnamefont
  {J.}~\bibnamefont {Ansermet}}, \ and\ \bibinfo {author} {\bibfnamefont
  {B.}~\bibnamefont {Barbara}},\ }\href {\doibase 10.1103/PhysRevLett.77.1873}
  {\bibfield  {journal} {\bibinfo  {journal} {Physical Review Letters}\
  }\textbf {\bibinfo {volume} {77}},\ \bibinfo {pages} {1873} (\bibinfo {year}
  {1996})}\BibitemShut {NoStop}%
\bibitem [{\citenamefont {Kent}\ \emph {et~al.}(1994)\citenamefont {Kent},
  \citenamefont {Moln{\'{a}}r}, \citenamefont {Gider},\ and\ \citenamefont
  {Awschalom}}]{Kent1994}%
  \BibitemOpen
  \bibfield  {author} {\bibinfo {author} {\bibfnamefont {A.~D.}\ \bibnamefont
  {Kent}}, \bibinfo {author} {\bibfnamefont {S.~V.}\ \bibnamefont
  {Moln{\'{a}}r}}, \bibinfo {author} {\bibfnamefont {S.}~\bibnamefont {Gider}},
  \ and\ \bibinfo {author} {\bibfnamefont {D.~D.}\ \bibnamefont {Awschalom}},\
  }\href {\doibase 10.1063/1.358160} {\bibfield  {journal} {\bibinfo  {journal}
  {Journal of Applied Physics}\ }\textbf {\bibinfo {volume} {76}},\ \bibinfo
  {pages} {6656} (\bibinfo {year} {1994})}\BibitemShut {NoStop}%
\bibitem [{\citenamefont {Geim}\ \emph {et~al.}(1997)\citenamefont {Geim},
  \citenamefont {Dubonos}, \citenamefont {Lok}, \citenamefont {Grigorieva},
  \citenamefont {Maan}, \citenamefont {Hansen},\ and\ \citenamefont
  {Lindelof}}]{Geim1997}%
  \BibitemOpen
  \bibfield  {author} {\bibinfo {author} {\bibfnamefont {A.~K.}\ \bibnamefont
  {Geim}}, \bibinfo {author} {\bibfnamefont {S.~V.}\ \bibnamefont {Dubonos}},
  \bibinfo {author} {\bibfnamefont {J.~G.~S.}\ \bibnamefont {Lok}}, \bibinfo
  {author} {\bibfnamefont {I.~V.}\ \bibnamefont {Grigorieva}}, \bibinfo
  {author} {\bibfnamefont {J.~C.}\ \bibnamefont {Maan}}, \bibinfo {author}
  {\bibfnamefont {L.~T.}\ \bibnamefont {Hansen}}, \ and\ \bibinfo {author}
  {\bibfnamefont {P.~E.}\ \bibnamefont {Lindelof}},\ }\href {\doibase
  10.1063/1.120034} {\bibfield  {journal} {\bibinfo  {journal} {Applied Physics
  Letters}\ }\textbf {\bibinfo {volume} {71}},\ \bibinfo {pages} {2379}
  (\bibinfo {year} {1997})}\BibitemShut {NoStop}%
\bibitem [{\citenamefont {Monzon}\ \emph {et~al.}(1997)\citenamefont {Monzon},
  \citenamefont {Johnson},\ and\ \citenamefont {Roukes}}]{Monzon1997}%
  \BibitemOpen
  \bibfield  {author} {\bibinfo {author} {\bibfnamefont {F.~G.}\ \bibnamefont
  {Monzon}}, \bibinfo {author} {\bibfnamefont {M.}~\bibnamefont {Johnson}}, \
  and\ \bibinfo {author} {\bibfnamefont {M.~L.}\ \bibnamefont {Roukes}},\
  }\href {\doibase 10.1063/1.120254} {\bibfield  {journal} {\bibinfo  {journal}
  {Applied Physics Letters}\ }\textbf {\bibinfo {volume} {71}},\ \bibinfo
  {pages} {3087} (\bibinfo {year} {1997})}\BibitemShut {NoStop}%
\bibitem [{\citenamefont {{Van Houten}}\ \emph {et~al.}(1986)\citenamefont
  {{Van Houten}}, \citenamefont {{Van Wees}}, \citenamefont {Heijman},\ and\
  \citenamefont {Andr{\'{e}}}}]{VanHouten1986}%
  \BibitemOpen
  \bibfield  {author} {\bibinfo {author} {\bibfnamefont {H.}~\bibnamefont {{Van
  Houten}}}, \bibinfo {author} {\bibfnamefont {B.~J.}\ \bibnamefont {{Van
  Wees}}}, \bibinfo {author} {\bibfnamefont {M.~G.~J.}\ \bibnamefont
  {Heijman}}, \ and\ \bibinfo {author} {\bibfnamefont {J.~P.}\ \bibnamefont
  {Andr{\'{e}}}},\ }\href {\doibase 10.1063/1.97243} {\bibfield  {journal}
  {\bibinfo  {journal} {Applied Physics Letters}\ }\textbf {\bibinfo {volume}
  {49}},\ \bibinfo {pages} {1781} (\bibinfo {year} {1986})}\BibitemShut
  {NoStop}%
\bibitem [{\citenamefont {Choi}\ \emph {et~al.}(1987)\citenamefont {Choi},
  \citenamefont {Tsui},\ and\ \citenamefont {Alavi}}]{Choi1987}%
  \BibitemOpen
  \bibfield  {author} {\bibinfo {author} {\bibfnamefont {K.~K.}\ \bibnamefont
  {Choi}}, \bibinfo {author} {\bibfnamefont {D.~C.}\ \bibnamefont {Tsui}}, \
  and\ \bibinfo {author} {\bibfnamefont {K.}~\bibnamefont {Alavi}},\ }\href
  {\doibase 10.1063/1.97869} {\bibfield  {journal} {\bibinfo  {journal}
  {Applied Physics Letters}\ }\textbf {\bibinfo {volume} {50}},\ \bibinfo
  {pages} {110} (\bibinfo {year} {1987})}\BibitemShut {NoStop}%
\bibitem [{\citenamefont {Heitmann}(1990)}]{Heitmann1990}%
  \BibitemOpen
  \bibfield  {author} {\bibinfo {author} {\bibfnamefont {D.}~\bibnamefont
  {Heitmann}},\ }in\ \href@noop {} {\emph {\bibinfo {booktitle} {Electronic
  Properties of multilayers and low-dimensional semiconductor structures}}},\
  \bibinfo {editor} {edited by\ \bibinfo {editor} {\bibfnamefont {J.~M.}\
  \bibnamefont {Chamberlain}}, \bibinfo {editor} {\bibfnamefont
  {L.}~\bibnamefont {Eaves}}, \ and\ \bibinfo {editor} {\bibfnamefont {J.-C.}\
  \bibnamefont {Portal}}}\ (\bibinfo  {publisher} {Springer},\ \bibinfo {year}
  {1990})\ pp.\ \bibinfo {pages} {151--173}\BibitemShut {NoStop}%
\bibitem [{\citenamefont {Monzon}\ \emph {et~al.}(1999)\citenamefont {Monzon},
  \citenamefont {Patterson},\ and\ \citenamefont {Roukes}}]{Monzon1999}%
  \BibitemOpen
  \bibfield  {author} {\bibinfo {author} {\bibfnamefont {F.~G.}\ \bibnamefont
  {Monzon}}, \bibinfo {author} {\bibfnamefont {D.~S.}\ \bibnamefont
  {Patterson}}, \ and\ \bibinfo {author} {\bibfnamefont {M.~L.}\ \bibnamefont
  {Roukes}},\ }\href {\doibase 10.1016/S0304-8853(98)01166-4} {\bibfield
  {journal} {\bibinfo  {journal} {Journal of Magnetism and Magnetic Materials}\
  }\textbf {\bibinfo {volume} {195}},\ \bibinfo {pages} {19} (\bibinfo {year}
  {1999})}\BibitemShut {NoStop}%
\bibitem [{\citenamefont {Grundler}\ \emph {et~al.}(1999)\citenamefont
  {Grundler}, \citenamefont {Meier}, \citenamefont {Broocks}, \citenamefont
  {Heyn},\ and\ \citenamefont {Heitmann}}]{Grundler1999}%
  \BibitemOpen
  \bibfield  {author} {\bibinfo {author} {\bibfnamefont {D.}~\bibnamefont
  {Grundler}}, \bibinfo {author} {\bibfnamefont {G.}~\bibnamefont {Meier}},
  \bibinfo {author} {\bibfnamefont {K.-B.}\ \bibnamefont {Broocks}}, \bibinfo
  {author} {\bibfnamefont {C.}~\bibnamefont {Heyn}}, \ and\ \bibinfo {author}
  {\bibfnamefont {D.}~\bibnamefont {Heitmann}},\ }\href {\doibase
  10.1063/1.370212} {\bibfield  {journal} {\bibinfo  {journal} {Journal of
  Applied Physics}\ }\textbf {\bibinfo {volume} {85}},\ \bibinfo {pages} {6175}
  (\bibinfo {year} {1999})}\BibitemShut {NoStop}%
\bibitem [{\citenamefont {Meier}\ \emph {et~al.}(2000)\citenamefont {Meier},
  \citenamefont {Grundler}, \citenamefont {Broocks}, \citenamefont {Heyn},\
  and\ \citenamefont {Heitmann}}]{Meier2000}%
  \BibitemOpen
  \bibfield  {author} {\bibinfo {author} {\bibfnamefont {G.}~\bibnamefont
  {Meier}}, \bibinfo {author} {\bibfnamefont {D.}~\bibnamefont {Grundler}},
  \bibinfo {author} {\bibfnamefont {K.-B.}\ \bibnamefont {Broocks}}, \bibinfo
  {author} {\bibfnamefont {C.}~\bibnamefont {Heyn}}, \ and\ \bibinfo {author}
  {\bibfnamefont {D.}~\bibnamefont {Heitmann}},\ }\href {\doibase
  10.1016/S0304-8853(99)00625-3} {\bibfield  {journal} {\bibinfo  {journal}
  {Journal of Magnetism and Magnetic Materials}\ }\textbf {\bibinfo {volume}
  {210}},\ \bibinfo {pages} {138} (\bibinfo {year} {2000})}\BibitemShut
  {NoStop}%
\bibitem [{\citenamefont {Schuh}\ \emph {et~al.}(2001)\citenamefont {Schuh},
  \citenamefont {Biberger}, \citenamefont {Bauer}, \citenamefont {Breuer},\
  and\ \citenamefont {Weiss}}]{Schuh2001}%
  \BibitemOpen
  \bibfield  {author} {\bibinfo {author} {\bibfnamefont {D.}~\bibnamefont
  {Schuh}}, \bibinfo {author} {\bibfnamefont {J.}~\bibnamefont {Biberger}},
  \bibinfo {author} {\bibfnamefont {A.}~\bibnamefont {Bauer}}, \bibinfo
  {author} {\bibfnamefont {W.}~\bibnamefont {Breuer}}, \ and\ \bibinfo {author}
  {\bibfnamefont {D.}~\bibnamefont {Weiss}},\ }\href {\doibase
  10.1109/20.951063} {\bibfield  {journal} {\bibinfo  {journal} {IEEE
  Transactions on Magnetics}\ }\textbf {\bibinfo {volume} {37}},\ \bibinfo
  {pages} {2091} (\bibinfo {year} {2001})}\BibitemShut {NoStop}%
\bibitem [{\citenamefont {Thornton}\ \emph {et~al.}(1986)\citenamefont
  {Thornton}, \citenamefont {Pepper}, \citenamefont {Ahmed}, \citenamefont
  {Andrews},\ and\ \citenamefont {Davies}}]{Thornton1986}%
  \BibitemOpen
  \bibfield  {author} {\bibinfo {author} {\bibfnamefont {T.~J.}\ \bibnamefont
  {Thornton}}, \bibinfo {author} {\bibfnamefont {M.}~\bibnamefont {Pepper}},
  \bibinfo {author} {\bibfnamefont {H.}~\bibnamefont {Ahmed}}, \bibinfo
  {author} {\bibfnamefont {D.}~\bibnamefont {Andrews}}, \ and\ \bibinfo
  {author} {\bibfnamefont {G.~J.}\ \bibnamefont {Davies}},\ }\href {\doibase
  10.1103/PhysRevLett.56.1198} {\bibfield  {journal} {\bibinfo  {journal}
  {Physical Review Letters}\ }\textbf {\bibinfo {volume} {56}},\ \bibinfo
  {pages} {1198} (\bibinfo {year} {1986})}\BibitemShut {NoStop}%
\bibitem [{\citenamefont {Zheng}\ \emph {et~al.}(1986)\citenamefont {Zheng},
  \citenamefont {Wei}, \citenamefont {Tsui},\ and\ \citenamefont
  {Weimann}}]{Zheng1986}%
  \BibitemOpen
  \bibfield  {author} {\bibinfo {author} {\bibfnamefont {H.~Z.}\ \bibnamefont
  {Zheng}}, \bibinfo {author} {\bibfnamefont {H.~P.}\ \bibnamefont {Wei}},
  \bibinfo {author} {\bibfnamefont {D.~C.}\ \bibnamefont {Tsui}}, \ and\
  \bibinfo {author} {\bibfnamefont {G.}~\bibnamefont {Weimann}},\ }\href
  {\doibase 10.1103/PhysRevB.34.5635} {\bibfield  {journal} {\bibinfo
  {journal} {Physical Review B}\ }\textbf {\bibinfo {volume} {34}},\ \bibinfo
  {pages} {5635} (\bibinfo {year} {1986})}\BibitemShut {NoStop}%
\bibitem [{\citenamefont {Ford}\ \emph {et~al.}(1988)\citenamefont {Ford},
  \citenamefont {Thornton}, \citenamefont {Newbury}, \citenamefont {Pepper},
  \citenamefont {Ahmed}, \citenamefont {Peacock}, \citenamefont {Ritchie},
  \citenamefont {Frost},\ and\ \citenamefont {Jones}}]{Ford1988}%
  \BibitemOpen
  \bibfield  {author} {\bibinfo {author} {\bibfnamefont {C.~J.~B.}\
  \bibnamefont {Ford}}, \bibinfo {author} {\bibfnamefont {T.~J.}\ \bibnamefont
  {Thornton}}, \bibinfo {author} {\bibfnamefont {R.}~\bibnamefont {Newbury}},
  \bibinfo {author} {\bibfnamefont {M.}~\bibnamefont {Pepper}}, \bibinfo
  {author} {\bibfnamefont {H.}~\bibnamefont {Ahmed}}, \bibinfo {author}
  {\bibfnamefont {D.~C.}\ \bibnamefont {Peacock}}, \bibinfo {author}
  {\bibfnamefont {D.~A.}\ \bibnamefont {Ritchie}}, \bibinfo {author}
  {\bibfnamefont {J.~E.~F.}\ \bibnamefont {Frost}}, \ and\ \bibinfo {author}
  {\bibfnamefont {G.~A.~C.}\ \bibnamefont {Jones}},\ }\href
  {http://journals.aps.org/prb/abstract/10.1103/PhysRevB.38.8518} {\bibfield
  {journal} {\bibinfo  {journal} {Physical Review B}\ }\textbf {\bibinfo
  {volume} {38}},\ \bibinfo {pages} {8518} (\bibinfo {year}
  {1988})}\BibitemShut {NoStop}%
\bibitem [{\citenamefont {Bureau-Oxton}\ \emph {et~al.}(2013)\citenamefont
  {Bureau-Oxton}, \citenamefont {{Camirand Lemyre}},\ and\ \citenamefont
  {Pioro-Ladri{\`{e}}re}}]{Bureau-Oxton2013}%
  \BibitemOpen
  \bibfield  {author} {\bibinfo {author} {\bibfnamefont {C.}~\bibnamefont
  {Bureau-Oxton}}, \bibinfo {author} {\bibfnamefont {J.}~\bibnamefont
  {{Camirand Lemyre}}}, \ and\ \bibinfo {author} {\bibfnamefont
  {M.}~\bibnamefont {Pioro-Ladri{\`{e}}re}},\ }\href
  {http://www.jove.com/video/50581/nanofabrication-of-gate-defined-gaasalgaas-%
lateral-quantum-dots} {\bibfield  {journal} {\bibinfo  {journal} {Journal of
  Visualized Experiments}\ }\textbf {\bibinfo {volume} {81}},\ \bibinfo {pages}
  {1} (\bibinfo {year} {2013})}\BibitemShut {NoStop}%
\bibitem [{\citenamefont {Birner}\ \emph {et~al.}(2007)\citenamefont {Birner},
  \citenamefont {Zibold}, \citenamefont {Andlauer}, \citenamefont {Kubis},
  \citenamefont {Sabathil}, \citenamefont {Trellakis},\ and\ \citenamefont
  {Vogl}}]{Birner2007}%
  \BibitemOpen
  \bibfield  {author} {\bibinfo {author} {\bibfnamefont {S.}~\bibnamefont
  {Birner}}, \bibinfo {author} {\bibfnamefont {T.}~\bibnamefont {Zibold}},
  \bibinfo {author} {\bibfnamefont {T.}~\bibnamefont {Andlauer}}, \bibinfo
  {author} {\bibfnamefont {T.}~\bibnamefont {Kubis}}, \bibinfo {author}
  {\bibfnamefont {M.}~\bibnamefont {Sabathil}}, \bibinfo {author}
  {\bibfnamefont {A.}~\bibnamefont {Trellakis}}, \ and\ \bibinfo {author}
  {\bibfnamefont {P.}~\bibnamefont {Vogl}},\ }\href {\doibase
  10.1109/TED.2007.902871} {\bibfield  {journal} {\bibinfo  {journal} {IEEE
  Transactions on Electron Devices}\ }\textbf {\bibinfo {volume} {54}},\
  \bibinfo {pages} {2137} (\bibinfo {year} {2007})}\BibitemShut {NoStop}%
\bibitem [{Note1()}]{Note1}%
  \BibitemOpen
  \bibinfo {note} {Mathematica Radia package available at http://www.esrf.eu/
  is used for the simulations of the stray magnetic field.}\BibitemShut {Stop}%
\bibitem [{\citenamefont {Klitzing}\ \emph {et~al.}(1980)\citenamefont
  {Klitzing}, \citenamefont {Dorda},\ and\ \citenamefont
  {Pepper}}]{Klitzing1980}%
  \BibitemOpen
  \bibfield  {author} {\bibinfo {author} {\bibfnamefont {K.~V.}\ \bibnamefont
  {Klitzing}}, \bibinfo {author} {\bibfnamefont {G.}~\bibnamefont {Dorda}}, \
  and\ \bibinfo {author} {\bibfnamefont {M.}~\bibnamefont {Pepper}},\ }\href
  {\doibase 10.1103/PhysRevLett.45.494} {\bibfield  {journal} {\bibinfo
  {journal} {Physical Review Letters}\ }\textbf {\bibinfo {volume} {45}},\
  \bibinfo {pages} {494} (\bibinfo {year} {1980})}\BibitemShut {NoStop}%
\bibitem [{\citenamefont {Roukes}\ \emph {et~al.}(1987)\citenamefont {Roukes},
  \citenamefont {Scherer}, \citenamefont {Allen}, \citenamefont {Craighead},
  \citenamefont {Ruthen}, \citenamefont {Beebe},\ and\ \citenamefont
  {Harbison}}]{Roukes1987}%
  \BibitemOpen
  \bibfield  {author} {\bibinfo {author} {\bibfnamefont {M.~L.}\ \bibnamefont
  {Roukes}}, \bibinfo {author} {\bibfnamefont {A.}~\bibnamefont {Scherer}},
  \bibinfo {author} {\bibfnamefont {S.~J.}\ \bibnamefont {Allen}}, \bibinfo
  {author} {\bibfnamefont {H.~G.}\ \bibnamefont {Craighead}}, \bibinfo {author}
  {\bibfnamefont {R.~M.}\ \bibnamefont {Ruthen}}, \bibinfo {author}
  {\bibfnamefont {E.~D.}\ \bibnamefont {Beebe}}, \ and\ \bibinfo {author}
  {\bibfnamefont {J.~P.}\ \bibnamefont {Harbison}},\ }\href
  {http://journals.aps.org/prl/abstract/10.1103/PhysRevLett.59.3011} {\bibfield
   {journal} {\bibinfo  {journal} {Physical Review Letters}\ }\textbf {\bibinfo
  {volume} {59}},\ \bibinfo {pages} {3011} (\bibinfo {year}
  {1987})}\BibitemShut {NoStop}%
\bibitem [{\citenamefont {Beenakker}\ and\ \citenamefont {van
  Houten}(1990)}]{Beenakker1990}%
  \BibitemOpen
  \bibfield  {author} {\bibinfo {author} {\bibfnamefont {C.~W.~J.}\
  \bibnamefont {Beenakker}}\ and\ \bibinfo {author} {\bibfnamefont
  {H.}~\bibnamefont {van Houten}},\ }in\ \href@noop {} {\emph {\bibinfo
  {booktitle} {Electronic Properties of multilayers and low-dimensional
  semiconductor structures}}},\ \bibinfo {editor} {edited by\ \bibinfo {editor}
  {\bibfnamefont {J.~M.}\ \bibnamefont {Chamberlain}}, \bibinfo {editor}
  {\bibfnamefont {L.}~\bibnamefont {Eaves}}, \ and\ \bibinfo {editor}
  {\bibfnamefont {J.-C.}\ \bibnamefont {Portal}}}\ (\bibinfo  {publisher}
  {Springer},\ \bibinfo {year} {1990})\ pp.\ \bibinfo {pages}
  {75--94}\BibitemShut {NoStop}%
\bibitem [{\citenamefont {Beenakker}\ and\ \citenamefont {van
  Houten}(1989)}]{Beenakker1989}%
  \BibitemOpen
  \bibfield  {author} {\bibinfo {author} {\bibfnamefont {C.~W.~J.}\
  \bibnamefont {Beenakker}}\ and\ \bibinfo {author} {\bibfnamefont
  {H.}~\bibnamefont {van Houten}},\ }\href
  {http://journals.aps.org/prl/abstract/10.1103/PhysRevLett.63.1857} {\bibfield
   {journal} {\bibinfo  {journal} {Physical Review Letters}\ }\textbf {\bibinfo
  {volume} {63}},\ \bibinfo {pages} {1857} (\bibinfo {year}
  {1989})}\BibitemShut {NoStop}%
\bibitem [{\citenamefont {Ebert}\ \emph {et~al.}(1983)\citenamefont {Ebert},
  \citenamefont {von Klitzing}, \citenamefont {Ploog},\ and\ \citenamefont
  {Weimann}}]{Ebert1983}%
  \BibitemOpen
  \bibfield  {author} {\bibinfo {author} {\bibfnamefont {G.}~\bibnamefont
  {Ebert}}, \bibinfo {author} {\bibfnamefont {K.}~\bibnamefont {von Klitzing}},
  \bibinfo {author} {\bibfnamefont {K.}~\bibnamefont {Ploog}}, \ and\ \bibinfo
  {author} {\bibfnamefont {G.}~\bibnamefont {Weimann}},\ }\href
  {http://iopscience.iop.org/article/10.1088/0022-3719/16/28/012/meta;jsession%
id=8BBAC3A60ADC085DC282F3360C2AE871.c3.iopscience.cld.iop.org} {\bibfield
  {journal} {\bibinfo  {journal} {Journal of Physics C}\ }\textbf {\bibinfo
  {volume} {16}},\ \bibinfo {pages} {5441} (\bibinfo {year}
  {1983})}\BibitemShut {NoStop}%
\bibitem [{Note2()}]{Note2}%
  \BibitemOpen
  \bibinfo {note} {SEM observation was performed on device C before
  measurements causing telegraphic noise and lowering the signal-to-noise
  ratio.}\BibitemShut {Stop}%
\end{thebibliography}%

\end{document}